A New Look at the Sub-electron Controversy

Of Milikan & Ehrenhaft

T. Datta, Ming Yin[1], Zhihua Cai and M. Bleiweiss[2]

NanoCenter and Physics & Astronomy Department

University of South Carolina

Columbia, SC 29208.

1) Permanent Address: Physics and Engineering Department

Benedict College, Columbia, SC 29204

2) Permanent Address: Physics Department,

Naval Academy Preparatory School

Newport, RI 02841

Abstract

The measurements of the electronic charge e, by Ehrenhaft and Millikan in the 1910's generated a bitter controversy between the "orthodox" and the atomic points of views. Guided by the Pauli-Fermi neutrino model for beta-decay in this paper we speculate that the introduction of a small hypothetical force is adequate to account for most of the contentious differences between Millikan's and Ehrenhaft's observations.

The electrical charge of an electron, e is one of the most precisely measured constants of nature. Currently, the committee on data for science and technology (CODATA) value of e is 1.602 176 53 x$10^{-19}$ C with a standard uncertainty of 14 in the last two digits[1]. It is an established fact that e is not an average value instead it is the quantum or the least amount of free electric charge. Also an object may only have a total charge exactly equal to integral multiples of e. Indeed the atomic nature of matter and the theory of quantum mechanics seem self evident in contemporary science.

Remarkably, the acceptance of quantization was far from automatic, challenge to the new paradigm continued well into the 1900's. Both the "orthodox and the heterodox" sides were confronted with the discoveries of atomic structure and radioactivity. Notables such as Born, Lorentz, Mach, Planck, Schrodinger, Sommerfeld and others were involved in the debate. At this time the pioneering measurements of e were in progress, its exact value and whether it has a minimum value were vigorously contested[2]. Confirmation of the quantization of e was one of the corner stones for the acceptance of the atomic theory and hence the "modern" perspective of physical sciences.

As mentioned above the question of the electric charge of an electron was very contentious. Two of the main protagonists in this controversy were R.A. Millikan[3] and F. Ehrenhaft[4]. Millikan[5] argued that his oil drop experiment provided direct proof for quantization. That is (i) there is a minimum unit of electric charge e, and (ii) the charge of any object occurs in whole number multiples of this unit. Quantitatively, expressed as histograms of number of observations vs the measured charge q, Millikan's graph is discreet and sharp, as can be seen in figure 1. In contrast Ehrenhaft using the best instrumentations in extant such as the ultra microscope refuted Millikan's ideas and introduced the idea of subelectronic charges[4]. Claiming better accounting of all the forces and a wider parameter range Ehrenhaft reported electrical charge on metal particulates often are smaller than e and deviates from multiplicity. His distribution is broad, the minimum value of the measured charge $q_{min}$ appears to be substantially smaller than e; also a significant part (shown grey) of counts correspond to sub-electronic charge and the width of each broad maxima is comparable to Millikan's unit step. However, in our opinion the sub-electron issue aside Ehrenhaft ought to have noticed the periodicity (quantization) in the charge distribution in his own data.

Curiously, in a number ways the differences in the Millikan and Ehrenhaft's histograms are paralleled by the spectra of alpha and beta decay. And not surprisingly during the early years of nuclear physics this difference also created considerable consternations. In this article we speculate if the Millikan-Ehrenhaft conundrum can be resolved a la' Fermi.

The spectra of Alpha decays are sharp and well defined with energy in the ~ MeV range, the typical nuclear energy scale. In contrast the main beta distributions, the continuous spectrum are broad and only a few decays attain million electron volts of energy. These differences confounded then current ideas about the validity of conservation laws of nature. The Pauli- Fermi neutrino hypothesis is the solution of the $\alpha-\beta$ paradox. Fermi[6] argued that in both types of decays conservation laws must remain valid. He postulated that Pauli's charge less neutral particle neutrino, must accompany the beta emission so that the mother nucleus shares momenta and energy with both the emitted electron (positron) as well as the neutrino. In this scenario, a third party the elusive neutrino caries away and hence obscures the energy (momentum) missing in the beta spectra. Fermi was courageous in bestowing such a critical role to the then undiscovered neutrino! The success of Fermi's theory was one of the confirmations of the quantum mechanical principles and impetuses that led to the eventual experimental observation of the neutrino.

In the case of e-measurement there was a heated controversy as to the correct determination of the particle mass and the response (force) of the surrounding medium (air). In practice both are determined from the balance of the weight force and the drag force at the terminal velocity of the particle under zero applied electric field. Each side debated the correct accounting of them. Millikan considered the drag to be a Stokes force calculated form $F_{Stoke} = \kappa V_{Ter}$ here the Stokes parameter $\kappa$ depends on the particle geometry as well as the terminal velocity ($V_{Ter}$) of the particle. It has been stated that Ehrenhaft was incorrect and a "failure" where as Millikan was a "success"[2]. However Ehrenhaft was a well-established physicist and employed sophisticated instrumentation and methodology[2]. For the present let us grant that each experimenter were employing the correct force appropriate to their respective particles. Hence, both could have been right. We will speculate on such a possibility in this paper.

Let us first consider Millikan's experiment in this case the forces on an oil drop of charge q and mass m is shown in figure 3a. The force balance between the downward weight and F$_{Stoke}$ with the upward electric force may be written as

$$qE = mg + F_{Stoke} \equiv mg(1 + \frac{V_E}{V_0})  \qquad 1a$$

or,

$$q = \frac{mg}{E}(1 + \frac{V_E}{V_0}) \qquad 1b$$

Where E is the applied electric field, V$_0$ is the terminal velocity under gravity (when E=0) and V$_E$ is the terminal velocity with E applied.

For Ehrenhaft's setup let us consider the scenario that in addition to the force due to the medium (F$_{drag}$) a (hypothetical) force X acts on the charged metal particle, as shown in figure 3b. Here it is postulated that X force arises only in presence of the applied field. With the X force present the vertical force balance equation would now become,

$$qE_x = mg + F_{drag} + X\cos\theta \qquad 2$$

In equation 2, θ is the angle between the electric force and the hypothetical force X and ranges between 0 and 180 degrees. The subscripts "x" refers to the values when X is present. However if one is unaware of the existence of X then one will be lead to believe that the charge on the particle is $q_x$. In other words, for the same particle mass and measured terminal velocities, one would insist as in equation 1,

$$q_x E_x = mg + F_{drag} \qquad 3$$

From 1 and 3 we obtain,

$$q_x = q - \frac{X \cos\theta}{E_x}$$

hence,

$$q - \frac{X}{E_x} \leq q_x \leq q + \frac{X}{E_x} \qquad 3a$$

If we call q as the Millikan value and $q_x$ to be that for Ehrenhaft then from equation 3a we estimate the smallest value of $q_x$ and $\Delta q_x$ the spread in $q_x$ to be,

$$q_x)_{min} = q - \frac{X}{E_x} \qquad 3b$$

and

$$\Delta q_x = 2\left(\frac{X}{E_x}\right) \qquad 3c$$

From the literature values of q and the minimum of $q_x$ the spread is

$$\Delta q_x \sim q \qquad \text{3d}$$

Also, the typical value of the hypothetical force is can be estimated as

$$X = \Delta q_x E \sim qE \qquad \text{4a}$$

For an Ehrenhaft particle of radius $\sim 10^{-8}$ m and density $\sim 10^4$ kg/m$^3$ the estimated mass is $\sim 10^{-20}$ kg. Hence assuming that the weight force (mg) is comparable to the applied electric force then for X one would estimate,

$$X \sim mg \sim 10^{-19} N \qquad \text{4b}$$

## Concluding remarks

Here we have speculated that in the e-measurements there may be an additional force X. The magnitude of X is different in the two experiments of interest. In one, it is negligibly small and in the other, its contributions cannot be ignored.

Millikan's particles were an order of magnitude larger in size than Ehrenhaft's, oils are typically non conducting dielectrics and liquid drops tend to be spherically shaped. Although in the 1900's it was impossible to resolve the geometry of submicron particles, as can be seen in the electro-micrograph (Figure 4) most likely metal particles Ehrenhaft's (Pt, Ag, Au, etc)were non-spherical in shape. These differences in electrical properties, size and shape may be the reason why X is a more critical factor for Ehrenhaft.

Conductivity is emphasized because X appears only when the electric field is non-zero. In addition it may also be of relevance to a more modern search for free quarks in niobium particles reported in "Evidence for the Existence of Fractional Charge on Matter"[7]. Compared with dielectric oil drops high conductivity metal particles will create far larger perturbations in the applied electric field distribution. If this force show up only under certain suitable conditions then such perturbations may go unnoticed but be sufficient to generate false estimates of charge. Because, large electric fields are not common in nature X may remain undetected if not specifically searched for. Needless to say at the minimum this is an extreme heretical idea but such forces can be of great relevance in many important nanoscale systems.

## Acknowledgements

We wish to thank Prof. Andrzej Staruszkiewicz for bringing this controversy to our attention. We thank the USC electron microscopy center for the picture of the nano-particulates of gold, shown in figure 4. This work is partially supported by funds from the USC Nanocenter.

Figures and Captions

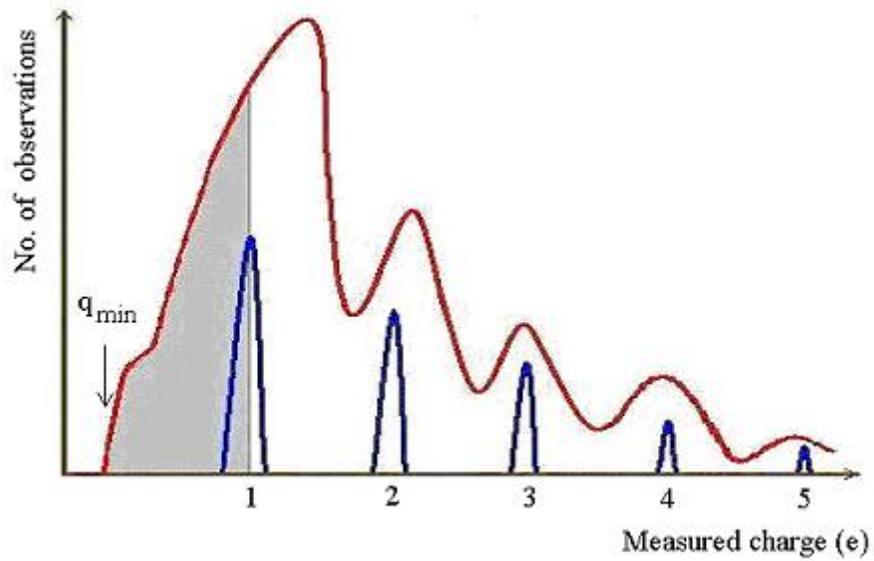

Figure 1: A semi-quantitative comparison of the competing charge histograms. Millikan's graph (blue) is sharp and discreet in equal steps of e. Ehrenhaft's curve is distribution (red) of broad overlapping peaks also the minimum value $q_{min}$ is much smaller than e.

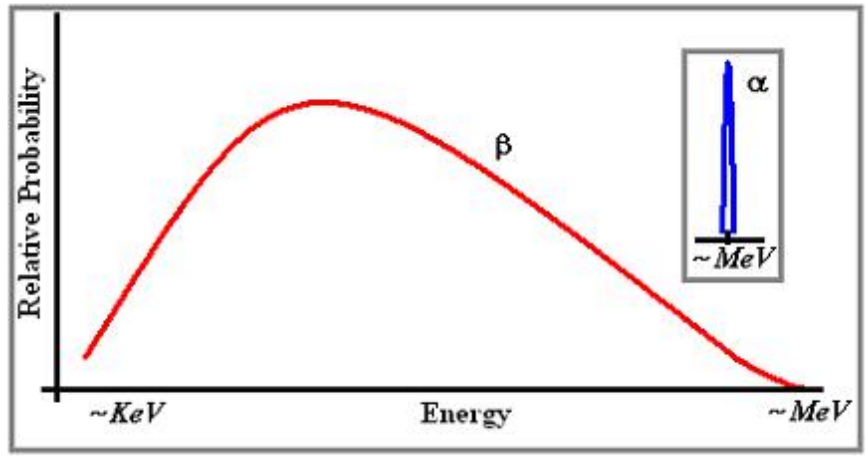

Figure 2: The distribution of alpha and beta particles produces in radioactive decays.

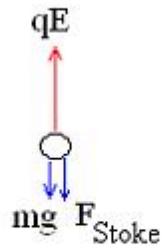

Figure 3a: The three forces on a charged Millikan oil drop.

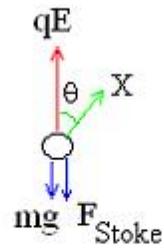

Figure 3b: A particle in the Ehrenhaft experiment with the additional X force acting on it.

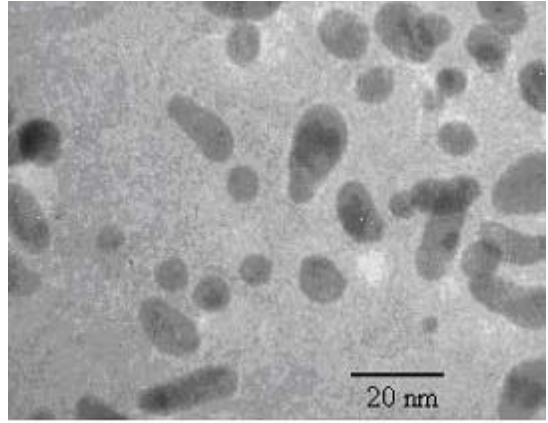

Figure 4: A scanning electron micrograph showing the non-spherical shapes of typical gold particulate in the 10 -100 nm range.